\documentclass[twocolumn,showpacs,secnumarabic,amssymb,nobibnotes,aps,prl]
{revtex4-1}

\usepackage[dvipdfm,bookmarks=true]{hyperref}
\usepackage{graphicx}
\usepackage{enumerate}

\begin{document}

\title{Topological One-Way Quantum Computation
on Verified Logical Cluster States}
\author{Keisuke Fujii}
\author{Katsuji Yamamoto}
\affiliation{
Department of Nuclear Engineering, Kyoto University, Kyoto 606-8501, Japan}
\date{\today}

\begin{abstract}
We present a scheme to improve the noise threshold for fault-tolerant
topological one-way computation with constant overhead.
Certain cluster states of finite size, say star clusters, are constructed
with logical qubits through an efficient verification process
to achieve high fidelity.
Then, the star clusters are connected near-deterministically
with verification to form a three-dimensional cluster state
to implement topological one-way computation.
The necessary postselection for verification is localized
within the star clusters, ensuring the scalability of computation.
By using the Steane seven-qubit code for the logical qubits, 
this scheme works with a high error rate of $ 2 \% $
and reasonable resources comparable to or less than
those for the other fault-tolerant schemes.
A higher noise threshold would be achieved by adopting a larger code.
\end{abstract}

\maketitle

{\it Introduction.---}
The issue of decoherence is one of the most important obstacles
for realization of quantum information processing.
To overcome this problem, fault-tolerant computation
based on quantum error correction (QEC) codes has been developed
\cite{Shor95,Calderbank-Shor96,Steane96,DiVincenzo-Shor96,Gottesman97}.
The main achievement of the quantum fault-tolerant theory
is the threshold theorem;
if the amount of noise per gate is smaller than a certain value,
namely the noise threshold, quantum computation can be performed
to arbitrary accuracy with a polynomial overhead \cite{threshold}.
The noise threshold has been calculated to be about $10^{-4}-10^{-2}$
for a variety of fault-tolerant schemes based on concatenated QEC codes
\cite{Steane03,Knill05,Aliferis,FY10FT}.
Besides this standard QEC method, there is a different promising approach
for fault-tolerance, where a surface code protects information
by virtue of topological degeneracy, without requiring concatenation
\cite{Kitaev97b}.
Then, one-way computation (OWC) \cite{OWC}
with the topological fault-tolerance can be performed
on a three-dimensional (3D) cluster state \cite{3D}.
Furthermore, the 3D system is mapped to a two-dimensional (2D) lattice
\cite{2D}.
The topological computation can be performed
only with nearest-neighbor two-qubit gates.
These 2D and 3D computations achieve noise thresholds
of $ 0.75 \% $ and $ 0.67 \% $, respectively.

On the other hand, if one needs to perform computations
by using noisy devices with an error rate of $ \sim 1 \% $,
some different approaches or additional ingredients will be required.
Here, we consider integrating the QEC encoding and postselection
into the topological one-way computation (TOWC) in a 3D cluster state.
In an early approach, improved preparation of encoded ancilla states
with postselection is considered \cite{Reichardt04,Knill05}.
Then, such an approach is applied to the OWC with offline preparation
of logical qubits and cluster states \cite{FY07,LC}.
The postselection to reduce the logical error efficiently, however,
appears to have trouble with scalable computation.
This dilemma between postselection and scalability
has been solved recently in a cluster-based architecture
by using the unique feature of OWC \cite{FY10FT}.

In this article, we present an efficient method
to construct arbitrarily large cluster states of logical qubits
with high fidelity, where postselection is adopted for verification
being reconciled with scalability.
Then, we apply this method to the TOWC in 3D cluster states
to improve its noise threshold.
That is, the TOWC is performed by using logical qubits,
where the logical degree of freedom is utilized
to reduce the logical measurement errors.
This is viewed as concatenation of the topological surface code
with a suitable QEC code.
The whole procedure consists of
(i) logical cluster-state preparation with verification,
(ii) near-deterministic connection with verification,
and (iii) TOWC by measuring the logical qubits.
At stage (i) a specific finite-size cluster state of logical qubits
is copiously prepared offline with postselection
through an efficient verification process based on syndrome extraction
\cite{FY10FT}.
At stage (ii), these cluster states are connected near-deterministically
with verification to scalably form a 3D cluster state of logical qubits.
This verification process removes the additional errors introduced
by the gate operation for the connection, keeping
the logical qubits in the 3D cluster state clean enough to implement the TOWC
below the threshold of the surface code at stage (iii).
Since the encoding and verification processes
require nonlocal two-qubit gates at the physical level,
the present scheme loses the good geometrical property
of the topological computation.
It will nevertheless be worth realizing quantum computation
with a high error rate of $ \sim 1 \% $ and a reasonable overhead.

\begin{figure}
\centering
\scalebox{0.4}{\includegraphics*[0cm,0.5cm][17cm,21.5cm]{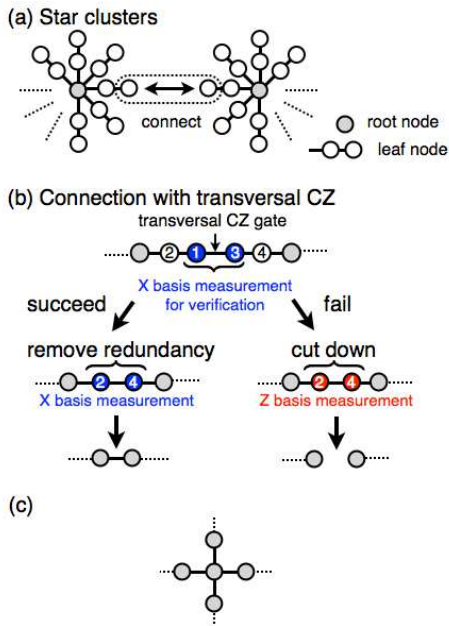}}
\caption{(a) Star clusters to be connected.
(b) Verification of the transversal C$Z$ gate
by measuring the qubits 1 and 3.
If the connection succeeds, the redundant qubits 2 and 4 are removed (left).
Otherwise, the connection is abandoned (right).
(c) By repeating this process, the root nodes are connected
to form a cluster state.}
\label{NDC}
\end{figure}

{\it Star clusters through double verification.---}
We can reduce the effective measurement error in OWC
by replacing each physical qubit with a logical one \cite{FY10FT,FY07,LC}.
It is, however, not a trivial task to prepare such large entangled states
as cluster states of logical qubits with high fidelity.
To this end a finite-size cluster state of logical qubits,
say a ``star cluster", is prepared via verification,
which consists of one ``root node" located at the center
and  $L$ surrounding ``leaf nodes," as shown in Fig. \ref{NDC} (a).

\begin{figure}
\centering
\scalebox{0.3}{\includegraphics*[0cm,1cm][18cm,7cm]{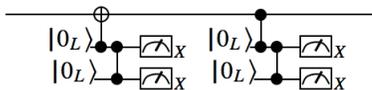}}
\caption{The double verification for the logical qubits
in the construction of a star cluster.}
\label{double}
\end{figure}

Starting with physical qubits, QEC code states $ |0_{L}\rangle $
and $|+_{L}\rangle =( |0_{L}\rangle + |1_{L}\rangle)/\sqrt{2}$
are first encoded by means of noisy gate operations as usual.
We adopt specifically the Steane seven-qubit code,
which is the minimum self-dual CSS (Calderbank-Shor-Steane) code
with distance three \cite{Calderbank-Shor96,Steane96}.
The logical qubits $ |+_{L}\rangle $'s are verified
by elaborately detecting the error syndrome
with primary and secondary ancilla qubits $ |0_{L}\rangle $'s
attached through transversal controlled-NOT (CNOT)
and controlled-$Z$ (C$Z$) gates, namely the double verification,
as shown in Fig. \ref{double}.
This double verification can optimally detect the first-order errors
\cite{FY09EP,FY10FT}.
The verified logical qubits are next connected with transversal C$Z$ gates
to form two-qubit logical cluster states.
The errors left on each qubit through the C$Z$ gate operations
are inspected further by the double verification.
Finally, $L$ verified two-qubit cluster states
and a single logical qubit are combined via transversal C$Z$ gates
with double verification to form a star cluster with $L$ leaf nodes.

Since all the verification procedures are transversal
on the seven-qubit code,
it is reasonably expected that the error distribution on each qubit
in the resultant star cluster is independent and identical,
(i.e., {\it homogeneous}) \cite{Eastin07}.
The errors on each qubit $ \rho $ at the physical level
are well approximated by $ \epsilon_A A \rho A $ ($ A = X,Y,Z $)
with probabilities
\begin{equation}
\epsilon_A = r_A p,
\label{eq:eA}
\end{equation}
which are determined in terms of the noise parameters
characterized by a mean error rate $ p $.
Here, we adopt the usual noise model:
(i) A two-qubit gate is followed by $A \otimes B$ errors
with probabilities $p_{AB}$ ($ A, B = I, X, Y, Z $, and $ AB \not= II $).
(ii) The preparation and measurement of physical qubits
are implemented with error probabilities $ p_P $ and $ p_M $, respectively.
Specifically, $ ( p_{AB}, p_P, p_M ) = (p/15,4p/15,4p/15) $
in the following analysis \cite{Knill05,FY10FT}.
Then, the leading-order calculation for double verification gives
$(\epsilon_X,\epsilon_Y,\epsilon_Z) = (p/15,p/15,2p/15)$,
which is in good agreement with the numerical result for $p \leq 2\%$
\cite{FY10FT,FY09EP}.
It is also checked for $p \leq 5\%$ that these errors are almost independent
among the physical-level qubits; the probabilities of correlated errors
are at least one order of magnitude smaller
than that of two simultaneous independent errors.

Given the homogeneous errors in Eq. (\ref{eq:eA}), the error probability
for the $X$-basis measurement of a single logical qubit
in the star cluster is calculated as
\begin{equation}
p_q^{(1)} \simeq f(p_{q}^{(0)})
\end{equation}
with the error probability $ p_q^{(0)} $ in measuring
each physical-level qubit in the $X$ basis as
\begin{equation}
p_q^{(0)} = \epsilon_Z + \epsilon_Y + p_M ,
\end{equation}
which is $ (7/15) p $ in the leading order with the double verification.
Specifically for the seven-qubit code,
$ p_q^{(1)} = 21 (p_{q}^{(0)})^2 + \ldots $
with $ f(x) = 1- (1-x)^7 - 7x(1-x)^6 $ .
The logical measurements in the $Z$ and $Y$ bases are less noisy
than that in the $X$ basis with $ \epsilon_Z > \epsilon_Y , \epsilon_X $
for double verification;
otherwise the modification of $ p_q^{(0)} $ is straightforward.

{\it Scalable construction of a 3D cluster state.---}
We can scalably construct a cluster state of an arbitrary size
by connecting the leaf nodes of the star clusters
with the transversal C$Z$ gates, as illustrated in Fig. \ref{NDC}.
Since additional errors are introduced by the C$Z$ gate operations,
they should be removed by a suitable verification,
which introduces the nondeterminism of postselection
to the connection process.
The situation is somewhat similar to the linear optical quantum computation,
where two-qubit gates are intrinsically non-deterministic
\cite{Nielsen04,Dawson06,Browne05Duan05}.
Thus, we follow the so-called {\it divide and conquer} approach
\cite{Nielsen04,Dawson06},
except that even if the connection has failed after all,
the cluster states are still connected erroneously
(in the case of the linear optical fusion gate,
the failure event results in a disconnected cluster state).

Specifically, in order to connect two neighboring root nodes,
the C$Z$ gates are operated between the ends of the two leaf nodes,
as shown in Fig. \ref{NDC} (b).
Then, the connected qubits 1 and 3 are measured in the logical $X$ basis
for verification.
If no errors are detected, the redundant qubits 2 and 4 are removed
by measuring them in the $X$-basis,
and the two root qubits are connected reliably.
On the other hand, if infection of error is found, the noisy connection
is abandoned by measuring the redundant qubits 2 and 4 in the $Z$ basis.
The success probability of this connection is estimated
for the seven-qubit code as $ p_{\rm s} \simeq
(1-p_G)^7 (1- \epsilon_X - \epsilon_Y - \epsilon_Z)^{14} (1- p_M)^{14} $.
Here, $ p_G = 4p/5 $ for the errors of the C$Z$ gate
except for $I \otimes X$, $X \otimes I$ and $X \otimes X$
commuting with the $X$-basis measurement.
By making several attempts, we can surely connect the root nodes,
as shown in Fig. \ref{NDC} (c).
If some or all of the connections have unfortunately failed
after consuming the $L$ leaf nodes,
the erroneous connections are used as though they have succeeded.
Such rare events can be included as the errors of logical qubits
with a probability $ p_{\rm fail} $, which will be calculated later.

\begin{figure}
\centering
\scalebox{0.3}{\includegraphics*[0cm,0.5cm][14cm,11.5cm]{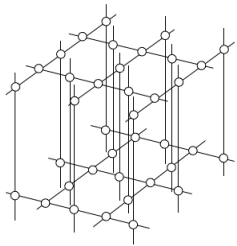}}
\caption{A 3D cluster state for TOWC, where 
each qubit is connected to four neighboring qubits
in a specific way \cite{3D}.
In the present scheme, this cluster state is constructed
by using the process in Fig. \ref{NDC}
so that each qubit is replaced by a logical one with high fidelity.}
\label{3DCluster}
\end{figure}

In this way, we can scalably connect the root nodes
to form an arbitrary cluster state of logical qubits
for fault-tolerant one-way computation.
Especially, the TOWC with a certain 3D cluster state
is promising to improve significantly the noise threshold
since the TOWC itself has a high noise threshold
with the surface code \cite{3D,2D}.
A 3D cluster for TOWC is shown in Fig. \ref{3DCluster},
where each qubit is connected with four neighboring qubits
in a specific way \cite{3D}.
In the present scheme, this cluster state is constructed
by using the process in Fig. \ref{NDC}
so that each qubit is replaced by logical one with high fidelity.

{\it Noise Threshold and resources.---} 
If the error probability $ q $ for measuring a single logical qubit
in the 3D cluster state is smaller
than the threshold value of the surface code,
the TOWC is performed in a fault-tolerant manner.
The noise threshold of the surface code with noisy syndrome measurements
has been obtained as 2.9 -- 3.3 \%
from the random plaquetta Z(2) gauge theory \cite{Wang03Ohno04}.
Thus, the threshold condition for the present scheme is given by
\begin{equation}
q (p) = p^{(1)}_q + L p^{(1)}_q + p_{\rm fail} < 3.3\% ,
\label{eq:q}
\end{equation}
where the logical error $ q (p) $ is given as a function
of the physical error $ p $, as seen so far.
The first and third terms, $ p^{(1)}_q $ and $ p_{\rm fail} $,
are responsible for the logical measurement errors of the root qubit
and the unsuccessful connection, respectively.
The second term $ L p^{(1)}_q $ comes from the Pauli by-products
which are introduced in removing the redundant qubits of the $L$ leaf nodes
by the $X$-basis (success case) or $Z$-basis (failure case) measurement.
The number of leaf nodes $L$ ($ > 4 $) is chosen to be sufficiently large,
so that the failure probability of the connection is reduced as
\begin{equation}
p_{\rm fail} = \sum_{k=0}^3
\left( \begin{array}{c} L \\ k \end{array} \right)
p_{\rm s}^k (1-p_{\rm s})^{L-k}
\sim \left( \begin{array}{c} L \\ 3 \end{array} \right) \times O(p^{L-3})
\end{equation}
with $ 1-p_{\rm s} \sim O(p) $,
where $ k $ represents the number of successful connections.
Here, it should be noted that the error probability of the $X$-basis
measurement conditioned on the correct syndrome for the verification,
as shown in the first procedure of Fig. \ref{NDC} (b), is estimated to be
of higher order as $ O (p^3) $ with the seven-qubit code.
This error thus provides a negligible contribution in Eq. (\ref{eq:q})
compared with $ p^{(1)}_q \sim O(p^2) $.
Numerically, for example, for the mean error rate $p = 1\% (2\%)$,
the success probability is $p_{\rm s} \simeq 0.88 (0.76)$
with the seven-qubit code.
The number of leaf nodes is chosen to be $L=7(9)$
so as to suppress the failure of connection
as $p_{\rm fail} \sim 0.1\%$.
Then, the error probability for the logical qubit becomes
$ q(p) = 0.98 \% (2.6 \%) $ for $p =1\% (2\%)$ in Eq. (\ref{eq:q}),
which is smaller than the threshold value of the surface code.
Universality can be obtained by using a noisy non-Clifford ancilla qubit
and the magic state distillation \cite{Bravyi05}.
The noise threshold for the magic state distillation
has been calculated to be at least 6.3\% \cite{2D}.
Therefore, we can conclude that the noise threshold is at least 2\%
in the present scheme for the fault-tolerant universal TOWC.

\begin{figure}
\centering
\scalebox{1}{\includegraphics*[1cm,1.75cm][12cm,6.75cm]{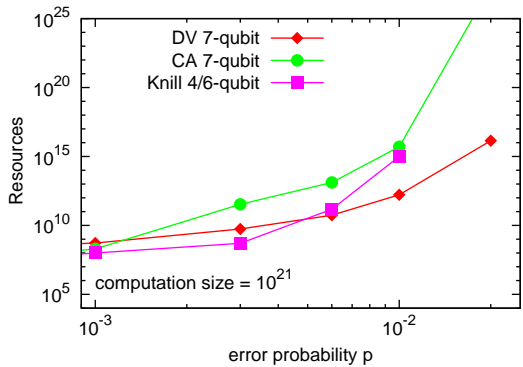}}
\caption{The resources per two-qubit gate for a computation size $10^{21}$
(gate accuracy $\sim 10^{-22}$) are plotted
as functions of the error rate $p$ for the present scheme
with double verification (red $ \Diamond $),
the cluster-based architecture (green $ \bigcirc $) \cite{FY10FT}
and the Knill's error-correcting architecture (purple $ \Box $)
\cite{Knill05}.}
\label{Resource}
\end{figure}

The resources per single two-qubit gate is calculated as $CR(q,\Omega)$
in terms of the resources for the TOWC
$R(q,\Omega)=[\ln (10\Omega)/\kappa (q)]^3 $
with $\kappa (q) \simeq (\ln 4q)/2$ \cite{3D}, where $q$ and $\Omega$
indicate the error probability and computation size, respectively.
The constant overhead $C$ for the logical encoding and verification
is given by
\begin{eqnarray}
C \sim N(1-  p)^{-K} ,
\end{eqnarray}
where $N$ is the number of physical qubits and gates per star cluster,
and $K$ is that of the error locations in the verification procedures,
respectively.
The resources are plotted as functions of $p$ in Fig. \ref{Resource}
for the present scheme with double verification
(DV 7-qubit, red $ \Diamond $) and the other competitive ones,
the cluster-based architecture (CA 7-qubit, green $ \bigcirc $)
\cite {FY10FT}
and the Knill's error-correcting architecture
(Knill 4/6-qubit, purple $ \Box $) \cite{Knill05}.
For example, in the present scheme  (DV 7-qubit),
we estimate for $ p = 1 \% $ and $L=7$ as $N \simeq 2 \times 10^4$,
$K \simeq 6 \times 10 ^2$ and $C \sim 6 \times 10^6$.
The overhead for the topological computation with $p = 1\% $
is given by $R(q = 0.98 \% , \Omega = 10^{21}) \sim 3  \times 10^5$.
Thus, the total overhead amounts to $CR \sim 2 \times 10^{12}$,
which is less by a few orders than those for the other schemes
operated with an error rate $ p = 1 \% $.

Although the cluster-based architecture suffers from memory errors
in the concatenation of QEC code (CA 7-qubit in Fig. \ref{Resource})
\cite{FY10FT}, the present scheme without several concatenations
does not accumulate the memory errors.
The memory errors are simply added as $p_{q}^{(0)} + \tau p \simeq
[ (7/15) + \tau ] p$, where $\tau$ denotes the effective waiting time.
The waiting time $\tau$ required for the memory is rather limited,
since we can run the TOWC on buffer nodes of finite size
with constructing the 3D cluster states in parallel.

Finally, we briefly discuss some possible improvements
for the performance of the present scheme.

We may start with preparing linear four-qubit logical cluster states
through the double verification, rather than two-qubit logical ones.
Then, as done in Ref. \cite{Chen06}, the star clusters are constructed
efficiently by connecting the four-qubit cluster states
with $X$-basis measurements for verification.
This reduces the overhead for the star-cluster construction
by virtue of parallelism, especially for an error rate
$ p \sim 2 \% $ or higher.

A larger QEC code such as the concatenated seven-qubit code
or the 23-qubit Golay code may be used for the logical qubits.
Then, the noise threshold of the TOWC would be further improved
since the double verification works even for $p=5\%$,
as indicated by numerical simulations
for the seven-qubit code \cite{FY10FT,FY09EP}.
For example, by assuming the homogeneous errors
we find that for the 23-qubit code the logical error probability $q(p)$
with $p=4\%$ is still below the TOWC threshold.
The total resources for $p=4\%$ amount to $CR \sim 2 \times 10^{30}$
with the resource efficient construction as discussed above.

A smaller QEC code may also be considered to save the overhead.
Specifically, by adopting the four-qubit error-detection code \cite{Knill05}
the logical error probabilities with $p=1\%$ are smaller
than the threshold values for TOWC \cite{Stace09}.
The resources per single logical gate with $p=1\%$
amount to $ 3 \times 10^{10}$,
in comparison with those for the DV 7-qubit case in Fig. \ref{Resource}.
However, with $p = 2\%$ the logical errors become
outside of the correctable region \cite{Stace09}.

\begin{acknowledgments}
This work was supported by JSPS Grant No. 20.2157.
\end{acknowledgments}

\end{document}